\documentclass[twocolumn,prl]{revtex4}
\RequirePackage{xspace}
\usepackage{graphicx}

\def\be{\begin{equation}}
\def\ee{\end{equation}}
\def\bearr{\begin{eqnarray}}
\def\eearr{\end{eqnarray}}
\def\tc{$T_c~$}

\def\c2{ CuO$_2~$}

\def\he4{${\rm {}^4He}$~}

\def\half{$\frac{1}{2}$~}

\begin{document}
\preprint{IMSc-2004/10/XX}

\title{Strongly Correlated Impurity Band Superconductivity in Diamond:\\
X-ray Spectroscopic evidence for upper Hubbard and mid-gap bands}  
  
\author{ G. Baskaran \\
Institute of Mathematical Sciences\\
C.I.T. Campus,
Chennai 600 113, India }

\begin{abstract}
In a recent X-ray absorption study in boron doped diamond, Nakamura et al.
have seen a well isolated narrow boron impurity band in non-superconducting 
samples and an additional narrow band at the chemical potential 
in a superconducting sample. We interpret the beautiful spectra as evidence 
for upper Hubbard band of a Mott insulating impurity band and an additional 
metallic `mid-gap band' of a conducting `self-doped' Mott insulator. This supports
the basic framework of a recent theory of the present author of strongly
correlated impurity band superconductivity (SCIBS) in a template of a wide-gap 
insulator, with no direct involvement of valence band states.
\end{abstract}
\maketitle

Recent discovery of superconductivity in boron doped diamond\cite{ekimov}
by Ekimov et al., is a pleasant surprise and superconductivity is seen in 
a place where one suspected it the least. This discovery has important 
implications in basic science and technology. These results have been 
confirmed\cite{takano,bustarret} at an elevated superconducting \tc in doped thin films synthesized 
by microwave plasma assisted chemical vapor deposition (MPCVD), a method which 
has advantages over the original high pressure-high temperature synthesis. 
This paves way for a series of new 
experiments\cite{bustarret,ekimov2,nakamura1,nakamura2} to unravel the mechanism 
of superconductivity in doped diamond as well as various 
anomalous properties that have been predicted theoretically\cite{dmndGB}. 
One such first measurements is a recent X-ray absorption spectroscopy 
(XAS) in the non-superconducting and superconducting dopings, by 
Nakamura et al\cite{nakamura1,nakamura2}.  

Unique among theoretical 
proposals\cite{dmndGB,boeri,lee,xiang,blase,russianTheory} for the mechanism 
of superconductivity
in this unusual system is a suggestion of the present author\cite{dmndGB}
that superconductivity
here is an impurity band phenomenon driven by strong electron correlation effects.
It has raised important questions on the origin and mechanism of superconductivity, 
by pointing out\cite{dmndGB} that the disordered superconducting state is 
in the vicinity of the Mott insulator-metal transition point\cite{shiomi,prinz}. 

In the XAS study in reference\cite{nakamura1,nakamura2}, a well isolated 
narrow boron impurity band is seen in non-superconducting samples and 
an additional narrow band at the chemical potential is seen in a 
superconducting sample. 
In this letter we interpret the spectra and provide evidence for an 
upper Hubbard band of an impurity band Mott insulator and an additional 
metallic `mid-gap band' of a conducting `self-doped' Mott insulator. 
This interpretation that is simple and natural supports the basic framework 
of a recent theory of the present author\cite{dmndGB} of superconductivity 
in a correlated impurity band in a template of a wide-gap insulator, 
with no direct involvement of valence band states, after the depletion of 
impurity states from the original broad valence band. We also discuss 
briefly at the end why an interpretation of the data in terms of certain 
lattice `relaxed' hole states may not be tenable.

On doping, a boron atom substitutes a carbon atom and bonds to 
neighboring carbon atoms through $sp^3$ hybridization. As boron has one less 
electron compared to carbon, a hole (an unpaired spin) is left behind, in an 
otherwise filled band. In the ground state this hole is bound to the boron 
atom in one of the three 
fold degenerate impurity state with a binding energy of 0.37 eV. As more and
more boron atoms are doped, the impurity wave functions overlap and there 
is an Anderson-Mott insulator to conductor transition at a critical 
boron concentration $n_c$, in this uncompensated semiconductor. The present 
author pointed out that since 
superconductivity observed in reference\cite{ekimov} is at a boron density, 
strikingly close to this critical density $n_c$\cite{shiomi,prinz,thonke}, 
superconductivity is likely to be 
an impurity band phenomenon, dominated by strong correlation physics within the
impurity band sub system. We will call this as the strongly correlated
impurity band 
superconductivity (SCIBS). We modeled SCIBS in terms of an effective 
repulsive Hubbard model and argued how resonating valence bond 
correlations\cite{pwascience} among spin-\half moments of neutral $B^0$ acceptor 
states and `self-doping', a spontaneous creation of a small and equal number of 
nominal $B^+$ and $B^-$ will lead to a Mott insulator to superconductor 
transition across a critical boron density $n_c$. In real systems there could 
be also compensation from donors such as nitrogen, phosphorus and
certain boron clusters.

In what follows we will interpret the recent XAS results as giving a strong
evidence for the presence of a correlated impurity band, an upper Hubbard   
band and creation of a `mid-gap band' by a process of self doping of the 
Mott insulator across the Mott insulator superconductor transition point.

In XAS, an electron from filled 1s state of either boron (B-K edge) or 
carbon (C-K edge)is removed and placed at an empty state above the 
chemical potential by
soft X-rays. We wish to probe the boron acceptor states close to the chemical 
potential. As boron has a strong $sp^3$ hybridization with carbon atoms, 
the impurity hole wave functions have a large radius of about 5 to 8 
Bohr radius. Thus it  has a large carbon wave function content. Further, 
in the range of concentration of interest acceptor wave functions start
overlapping so that XAS on C-K edge are good enough 
to bring out the nature of states close to the chemical potential. It is 
clear from the experiments that C-K edge spectra has an advantage 
over B-K edge, as it has less contamination from deep levels from 
an unavoidable density of interstitial boron atoms and boron clusters.

\begin{figure}
\includegraphics[width=9cm]{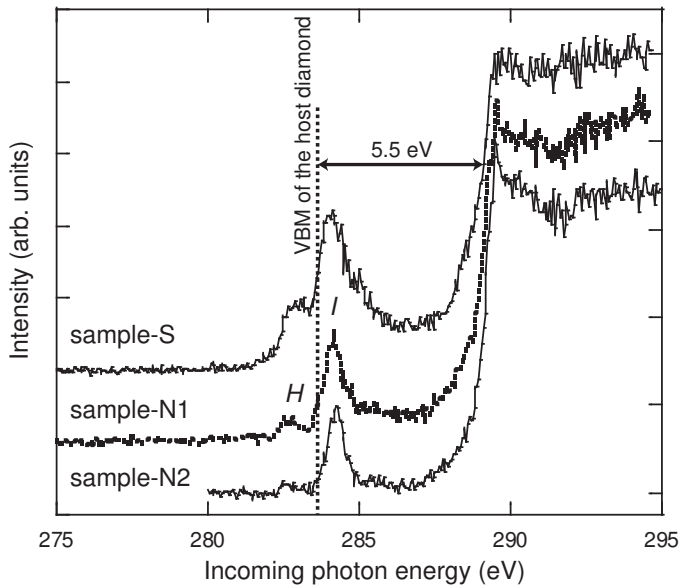}
\caption{\label{Fig1}
The C-$K$ XAS spectra of Nakamura et al.\cite{nakamura1} is reproduced;
S-superconducting sample and N1,N2 non-superconducting samples.
The dotted line represents the energy of the valence band maximum using 
the band gap of 5.5 eV.  Two characteristic peaks, $H$ and $I$, are observed 
at about 282.6 and 284.1eV. We interpret peak $I$ as upper Hubbard band and 
$H$ as the conducting mid-gap band of our impurity band subsystem (see text). 
Chemical potential lies within the small peak $H$, the metallic mid-gap band.}
\end{figure}
Before we proceed, we would like to make two observations
on the remarkable X-ray absorption spectra of reference\cite{nakamura1}: 
i) for all three boron concentrations where experiments
have been done, two non superconducting samples (below $n_c$) and one
superconducting sample (above $n_c$), the impurity band retains its integrity
and its center does not move. ii) at the chemical potential a new narrow
band evolves and gets prominent in the superconducting state and the chemical
potential is, surprisingly, pinned at an energy below the top of the valence band. 
\begin{figure}
\includegraphics[width=9cm]{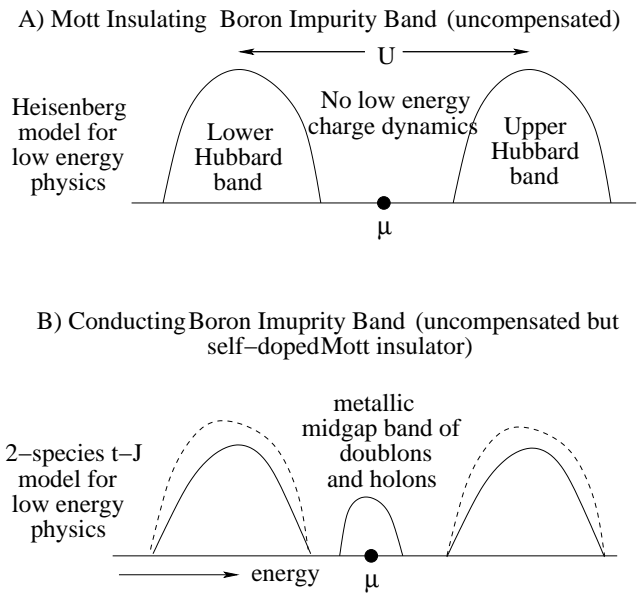}
\caption{\label{Fig2}
Schematic local single particle spectral function of an electron in impurity
band for two cases: A) Boron impurity band Mott insulator exhibiting upper 
and lower Hubbard bands and B) self-doped case, obtained by increasing boron 
density beyond Mott transition point; here mid-gap band contains equal number
of holons and doublons, whose density depends on band parameters and long 
range coulomb interaction. Hubbard band splitting by U and chemical potential 
$\mu$ is also shown. In describing low energy physics, elimination of upper 
and lower Hubbards bands generates superexchange and we obtain a Heisenberg 
model for case A and a 2 species t-J model\cite{rvbmsGB} for case B.
A standard spectral sum rule keeps the total area under upper, lower 
and mid-gap bands the same in both cases.}
\end{figure}

In what follows we discuss a simple theory of the shape of X-ray absorption
and emission spectra of impurity band Mott insulator and conductor in the 
vicinity of the metal insulator transition point.

In XAS, the absorption
cross section is determined, using Fermi golden rule and impulse approximation
as
\be
I_{\rm a}(\omega) \sim  \frac{2\pi}{\hbar} \sum_{f} 
| \langle i | \sum_{\ell} \frac{e}{c}{\bf p}_{\ell}\cdot {\bf A} | f \rangle  |^2 
\delta (E_f - E_i - \hbar \omega)
\ee

We are interested in a narrow range of energy just above the chemical
potential $\mu$ and ignore frequency dependence of  matrix elements  
to get 
\be
I_{\rm a}(\omega) ~~\propto~~ \rho_{\rm e}(\omega), ~~~ \hbar \omega~ >~ \mu
\ee
where $\rho_{\rm e}(\omega)$ is the local(atomic) one electron spectral 
function. The
X-ray fluorescence or emission (XES) spectra, where electrons close to 
the chemical potential fall into the empty 1s-core hole state and emit X-rays,
the emission intensity is given by
\be
I_{\rm e}(\omega) ~~\propto~~ \rho_{\rm h}(\omega), ~~~ \hbar \omega~ <~ \mu
\ee 
Here $\rho_{\rm e}(\omega)$ is the local one hole spectral function.

Now we discuss the profile of the electron and hole spectral functions.
At the energy scale of the impurity band 
width our discussion is rigorous and the spectral profile is qualitatively
correct. These spectral features of a Mott insulator and doped Mott insulator
are known from early Hubbard I to III approximations and recent 
infinite dimensional and dynamical mean field theory approaches\cite{infU}. 

In the atomic limit, $ U >> t_{ij}$, for a Hubbard model at half filling,
\be
\rho_{\rm e,h}(\omega) \sim~  \delta \left(\hbar \omega \pm 
\frac{U}{2}\right)
\ee
For our impurity Hubbard bands, disorder and finite $t_{ij}$ will broaden
the spectral function to give a shape shown in figure 2A. If we dope
the `half filled' impurity band with electron (doublons), for example by a 
small amount of compensation by an additional nitrogen doping, we introduce 
a band of mid-gap states, whose area is proportional to the
compensation (dopant) fraction $x$. Mid gap states of a paramagnetic state
are correlated many body energy levels and have no single particle description, 
for example in a Hartree-Fock theory. The low energy dynamics of the metallic 
band of mid-gap states (doublons) in this case of partial compensation by
nitrogen is governed by  t-J model. This is because, when we eliminate the 
surviving upper and lower Hubbard bands of a doped Mott insulator 
to obtain low energy effective Hamiltonian, superexchange gets generated. 

\begin{figure}
\includegraphics[width=9cm]{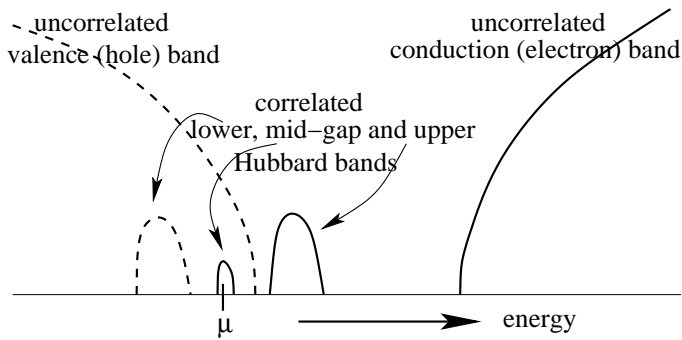}
\caption{\label{Fig3} We have overlaid the correlated Hubbard and mid-gap 
impurity bands with the one particle valence and conduction band density of 
states. Location of the chemical potential $\mu$ has meaning only within 
the Hubbard and mid-gap bands. As far the valence band it has no holes. 
XAS measures the conduction band, the upper Hubbard band and the 
mid-gap bands shown as thick lines. }
\end{figure}
In a recent paper we suggested\cite{rvbmsGB}, based on experimental results 
and theoretical
considerations on Mott insulator to metal transition, that the conducting
side close to the first order Mott transition point should be viewed as 
a `self doped' Mott insulator. As the doping is internal, number of 
negative and positive (doublons and holons) carriers are the same. The carrier
density is determined by long range coulomb interactions and band parameters.
To describe the low energy dynamics of such a doped Mott insulator, which 
preserves superexchange in the conducting state, we introduced a 2 species
t-J (or 2t-J) model. The schematic spectral profile, figure 2B, is the same 
as regular t-J model except that the mid-gap metallic band contains equal 
number of positive (holon) and negative (doublon) carriers (with reference 
to  neutral Mott insulator). 
As mentioned earlier, in reality there can be a small amount
of compensation from impurities such as nitrogen, phosphorus and 
certain boron clusters. The resonating valence bond physics of 
superconductivity in our self-doped Mott insulator\cite{rvbmsGB} is partly 
contained in the t-J-U model\cite{tJU}, but with key differences.

X-ray absorption spectroscopy measures the upper Hubbard band and 
the conducting `mid-gap band'. X-ray emission spectroscopy measures
the lower Hubbard band the conducting `mid-gap band'. However, in the
case of acceptor impurity band, such as ours, emission from the
close by valence band will mask the small signal from the impurity bands,
making XES problematic. In the same vein, if we had a donor band
XAS, rather than XES will have problems.

Now we discuss the experimental result of reference\cite{nakamura1}, 
figure 1, in the
light of  the above discussion. The spectra are for three boron dopings, two 
non-superconducting and one superconducting cases. The conduction band edge 
and the impurity bands are clearly visible. In the non-superconducting and
superconducting samples the impurity band remains sharp, denoted by the
peak $I$. This is our upper Hubbard band. In the superconducting state we 
have the additional band, at the chemical potential, which we identify as the 
metallic mid-gap band of self doped carriers, denoted by peak $H$. 
From the experimental result, we see that the 
peak to peak distance of the upper Hubbard band and mid-gap band,
$\frac{U}{2} \approx$ 0.7 to 1 eV. This value of U $\sim$ 1 to 2 eV
for our impurity orbital is easily rationalized, as the impurity Bohr
radius of our deep level acceptor state is increased only by a factor of 
5 to 8 from atomic Bohr radius.

Since U $\sim 1~eV$ is large compared to acceptor binding energy 
$E_B \approx 0.37~eV$, an important question in the conducting state about 
possible hole transfer to top of the valence band arises. That is, soon
after metallization increased charge fluctuations in the impurity band
will introduce the Hubbard U repulsion energy among the holes by an 
amount $\sim x~U$ prompting hole transfer to the valence band. 
Here $x$ density of self doped carriers. In addition to this energy loss 
on metallization, there are two sources of energy gain; i) the well known 
delocalization energy, proportional to the impurity band width and ii) 
a Madelung energy gain\cite{rvbmsGB} within the impurity band arising from 
long range 
coulomb interaction $E_{\rm Md}$. This energy is missed in a Hubbard model,
which ignores long range coulomb interaction; it is this which also 
determines the self doping fraction $x$.  When $ xU - E_{\rm Md} > E_B$, 
it will be energetically favorable to drain holes from the impurity band 
to the top of the valence band. 

In the experiments\cite{nakamura1}, the chemical potential neither falls on 
top of
the valence band nor inside the impurity band, but seems to be pinned to
the metallic mid-gap band. This means all low energy charge and spin actions 
are taking place within the impurity band. This is a good support for our 
impurity band modeling\cite{dmndGB} of the superconductivity phenomena. 
One may safely say that there are no hole fermi surface from a finite density 
of valence band holes, as is assumed in current electron-phonon 
theories\cite{boeri,lee,xiang,blase}.

At this point it is also important to make a qualitative difference
between a renormalized one particle band such as the conduction or 
valence band and a correlated many body bands such as upper, lower and
mid-gap Hubbard bands. A finite value of spectral functions in these many 
body bands are basically projected one particle density of states rather 
than one particle level densities. We emphasize this in figure 3, where we 
overlay the correlated Hubbard and mid-gap impurity bands on the 
one particle valence and conduction band density of states. Location
of the chemical potential $\mu$ has meaning only within the Hubbard and 
mid-gap bands. As far the valence band it has no holes. 
This figure also shows the mid-gap band, upper Hubbard band and the
conduction band that will be seen in the X-ray absorption spectra by a 
thick line. 
 
The authors of the XAS experiments\cite{nakamura1} suggest that what we call 
as the mid-gap
band at the chemical potential could arise from certain lattice relaxed hole
states in the absence of simple rigid band shifts on doping. Unfortunately
the energy of the hole is shifted in the opposite direction and it does not
look like a relaxed state but a `strained metastable state' which could 
decay down to the top of the valence band (energy of a hole is always 
positive and measured from the chemical potential). How, on metallization such
a meta stable hole states will collectivized and become stable ? That too in 
the background of an impurity band that continues to be present, as the XAS
spectra indicates. One also has to answer why relative density of such lattice 
relaxed hole states (area under mid-gap band relative to impurity band) should 
increase with boron doping.    

It is also not clear whether a phase separation of boron impurities will
explain the observed XAS spectra. Indeed, some careful analysis has been 
done in the experiments\cite{nakamura1} to eliminate the possibility of 
precipitation.

In all our discussion we did not discuss disorder explicitly. As emphasized
in our original paper, the uncompensated case under discussion has a 
commensurate filling of the impurity band, with an average of one hole per 
boron atom. In such a situation
Mott-Hubbard correlations overwhelm; disorder to a large extent is 
`irrelevant' in the renormalization group sense. That is, once Mott localization 
sets in, Anderson localization has lesser role to play. Further, spin charge 
decoupling, an inevitable consequence in the Mott transition region,
also protects current carrying holon and doublon states from Anderson localization 
effects. As we move far away from commensurate situation,
for example by partial compensation, Anderson localization effects and notions
such as mobility edge become important.

To summarize, it is indeed nice to have diamond, a relatively `simple' 
system as a template for the fascinating impurity band Mott phenomena and 
superconductivity. A large band gap and a relatively large acceptor binding 
energy of a simple acceptor boron, nicely isolates out and makes the impurity
Hubbard bands and mid-gap bands visible in X-ray absorption spectroscopy. 
It will be very nice to increase the energy resolution and explore some
of the beautiful many body issues related to our mechanism\cite{dmndGB} of 
SCIBS by X-ray absorption and emission spectroscopy, for a range of boron doping
and also partial compensation.

I thank Manas Sardar for useful discussions and bringing to my attention 
reference\cite{nakamura2}.

\end{document}